\documentclass[useAMS,usenatbib,12pt,epsfig,letter,multirow]{mn2e}
\usepackage[english]{babel}
\usepackage{multirow}
\usepackage{amssymb}
\usepackage{graphicx}

\title
[Cosmic dust and cosmology with supernovae]
{On the impact of intergalactic dust on cosmology with type Ia supernovae}
\author[M\'enard et al.]
{
\parbox[h]{\textwidth}{
Brice M\'enard$^{1}$, 
Martin Kilbinger$^2$,
Ryan Scranton$^3$} 
\vspace*{2pt} \\
\hspace{-.1cm}$^1$ Canadian Institute for Theoretical Astrophysics\\
\hspace{-.1cm}$^2$ Institut d'Astrophysique de Paris\\
\hspace{-.1cm}$^3$ University of California-Davis\\
}

\def\be{\begin{equation}}
\def\ee{\end{equation}}

\begin{document}

\date{Draft, \today}

\maketitle

\begin{abstract}
Supernova measurements have become a key ingredient in current determinations 
of cosmological parameters. These sources can however be used as standard candles only 
after correcting their apparent brightness for a number of effects. In this 
paper we discuss some limitations imposed by the formalism currently used for 
such corrections and investigate the impact on cosmological constraints. 
We show that color corrections are, in general, expected to be biased. In 
addition, color excesses which do not add a significant scatter to the
observed SN brightnesses affect the value of cosmological parameters but leave
the slope of the color-luminosity relation unchanged.

We quantify these biases in the context of the redshift-dependent dust 
extinction suggested by the recent detection of intergalactic dust 
by~\cite{Menard+09}.  Using a range of models for the opacity of the Universe 
as a function of redshift, we find that color-magnitude-stretch scaling 
relations are virtually insensitive to the presence of cosmic dust while 
cosmological parameters such as $\Omega_{\rm M}$ and $w$ are biased at the level 
of a few percent, i.e. offsets comparable to the current statistical errors.

Future surveys will be able to limit the impact of intergalactic extinction by 
observing at larger wavelengths. In addition such datasets will provide direct 
detections of intergalactic dust by cross-correlating SN colors and the 
density of foreground galaxies, which can be used as a consistency check on 
the cosmic dust extinction correction. Alternatively, such biases could be 
avoided by correcting the colors of supernovae on an object-by-object basis 
with accurate photometry.
\end{abstract}

\vspace{1cm}
\begin{keywords} 
supernovae -- cosmology: parameters -- dust: intergalactic
\end{keywords}

% =======================================================
\section{INTRODUCTION}
\label{sec:introduction}

About a decade ago, the use of type Ia supernovae (SNe Ia) as standardizable
candles led to the discovery of the accelerating expansion of the Universe 
(\citealt{1998ApJ...493L..53G,1998Natur.391...51P,1998AJ....116.1009R,1998ApJ...507...46S}). Since 
then, SNe Ia measurements have become a cornerstone in cosmological parameters 
estimation.  More recent and upcoming surveys are aiming to push these
constraints to higher accuracies and explore the properties of dark energy. 
To this end, it is important to identify and characterize the potential limits and biases involved.

The use of SNe Ia as standard candles requires correcting their observed
magnitudes for a number of effects: a luminosity--light curve shape dependence
(usually referred to as the ``stretch''), a luminosity--color relation, dust extinction and
gravitational magnification.  For sufficiently large surveys, magnification effects
average out, adding only scatter to the apparent magnitudes without
introducing any significant bias.  Conversely, the effect of dust extinction 
is more critical, as it is cumulative along the line of sight. In this
paper, we address the limits and potential biases of the procedure 
currently used to correct for such effects, focusing on the extinction
induced by intergalactic dust.

The presence of dust on large scales around galaxies has been the subject of 
numerous papers.  Theoretical studies indicate that dust grains can be
efficiently transported from galactic disks to the intergalactic medium 
through winds and radiation 
pressure~\citep{1999ApJ...525..583A,2001ApJ...561..521A,2005MNRAS.358..379B}.
Based on estimates of the stellar density and metallicity as a function of 
redshift, several authors have inferred the existence of significant amounts 
of intergalactic dust and a cosmic dust density
$\Omega_{\rm dust} \sim 10^{-6}-10^{-5}$~\citep{1997ApJ...490..571L,2007NewAR..51..332C,2004MNRAS.350..729I}.
The amount of dust in galaxy clusters has been explored through reddening measurements of background
sources (e.g. \citealt{2007ApJ...671L..97C,2008ApJ...688..198B,2008ApJ...680..975M}).
Upper limits on 
the the cosmic opacity were recently obtained by 
measuring the excess scatter seen in higher redshift quasar colors,  
\citep{2003JCAP...09..009M}, using the Toman test \citep{2008arXiv0810.5553M} and
combining constraints from luminosity distances and  ${\rm H}(z)$ 
\citep{2009arXiv0902.2006A} and using the amount of dust X-ray 
scattering around AGNs \citep{2009arXiv0902.4703D}.

Recently, \cite{Menard+09} (M09 hereafter) cross-correlated the colors of 
about 85,000 distant quasars with the position of 20 million $z\sim0.3$ 
galaxies observed with the Sloan Digital Sky Survey (\citealt{york2000}; 
SDSS).  They reported a statistical detection of dust reddening up to large 
scales around galaxies.  From this result they inferred the opacity of the 
Universe, calibrated at $z \sim 0.3$ and extrapolated to higher redshifts. 
Their model-dependent estimate gives $A_B(z=1)\sim\;{\rm a~few}\,\times 10^{-2}$ 
mag. Such an opacity is not negligible given the precision level of current 
supernovae surveys.  This Letter makes use of these results and addresses 
their impact on the estimation of cosmological parameters from SNe Ia.  
Our analysis does not include any effects due to a possible gray dust 
component. If present, the biases in
cosmological parameters derived below should be seen as lower limits only.

% =======================================================
\section{Cosmological information from type Ia supernovae}
\label{sec:cosmological_information}

% =======================================================
\subsection{Scaling relations}
\label{sec:scaling_relations}

The distance modulus $\mu$ of a standard candle with intrinsic absolute
magnitude $M$ can be used as an estimate of the luminosity distance:
\begin{equation}
\mu\equiv m-M=5\,\log_{10}\left(D_{\rm L}(z)/10\;{\rm pc}\right)\;,
\label{eq:mu}
\end{equation}
where $m$ is the observed magnitude.  The luminosity distance $D_{\rm L}$ is 
defined by the relationship between bolometric flux $S$ and luminosity $L$:
$D_{\rm L}\equiv \sqrt{L/(4\pi S)}$.  $D_{\rm L}$ can be related to the
cosmology by
\begin{equation}
D_{\rm L}(z)=\frac{c}{H_0}\,(1+z)\,\int_0^z\;\frac{\rm d z'}{E(z')}
\end{equation}
where, for a Universe containing a dark-energy component with density
$\Omega_{\rm X}$ and equation of state $w$,
\begin{equation}
E(z)\equiv \sqrt{\Omega_{\rm M}\,(1+z)^3+\Omega_{\rm X}(1+z)^{3(1+w)}}\;.
\end{equation}
The impact of dust extinction on the estimation of cosmological parameters 
using the luminosity distance can be illustrated by Taylor-expanding the
distance modulus from Equation~\ref{eq:mu}.  In the case of a flat Universe
with a cosmological constant, we find
\begin{equation}
\delta m \simeq \{-0.8,-1.3\} \;\delta \Omega_{\rm M} \mbox{~~~for a source at $z=\{0.5,1\}$}\;.
\label{eq:omega_m}
\end{equation}
Therefore, the \emph{mean} magnitude of standard candles must be controlled at
a level comparable to the targeted accuracy of the matter density estimation.
The above equations show that, unless corrected, the existence of a cosmic
opacity of $A_B\sim0.05$ will affect the value of $\Omega_{\rm M}$ by a similar 
amount, translating into a $\sim15\%$ bias for $\Omega_{\rm M}=0.3$.  As 
derived by \cite{2008ApJ...682..721Z}, a similar relationship can be written
down for the shift in the equation of state parameter $w$:
\begin{equation}
\label{eq:w}
\delta m\simeq 0.5\;\delta w\;.
\end{equation}

% =======================================================
\subsection{Luminosity distance and potential biases}
\label{sec:luminosity_distance}

Given a sample of supernovae, constraints on cosmological parameters can be
computed by comparing the expected and observed luminosity distances. To do so
we define the $\chi^2$-function,
\begin{equation}
  \chi^2 =  \sum_i \frac{\left[ \mu_i-5\,\log_{10}(D_{\rm L}(z_i)/10\;{\rm pc})
      \right]^2}{\sigma_i^2}\;.
  \label{eq:chi2}
\end{equation}
For each supernova, the distance modulus $\mu_i$ is calculated according to 
Equation~\ref{eq:mu} for a magnitude
\begin{equation}
m_i = m_{{\rm obs},i}-\delta m_i
\label{eq:sn}
\end{equation}
where $m_{{\rm obs},i}$ is the observed magnitude (usually expressed in terms of
the rest-frame $B$-band) and $\delta m_i$ is the net magnitude shift needed to
convert the supernova from a standardizable to a standard candle. 
Recent analyses have included two components to describe $\delta m$:
\begin{enumerate}
\item a correction term to account for any color-magnitude relation, including
  intrinsic color excess, dust extinction and reddening, etc.
\item a stretch correction to include an intrinsic dependence between the
  brightness and duration of SNe Ia 
\end{enumerate}
The latter term is not of interest in the present analysis; it will be 
included in all calculations but not discussed explicitly.

It is generally assumed that these correction terms are \emph{not} redshift 
dependent. This allows one to write the distance modulus as
\begin{equation}
  \mu_i=m_{{\rm obs},i}-M+\alpha(s_i-1)-\beta\,c_i\;.
\label{eq:mu_beta}
\end{equation}
where $\alpha(s_i-1)$ refers to the stretch correction, $\beta\;c_i$ is 
the color-based correction, making use of the observed color excess $c_i$ 
(usually expressed as the rest-frame E$(B-V)$ color) of each
supernova\footnote{Other teams have used a different parameterization to apply 
a color-based correction (\citealt{2007ApJ...659..122J,2007ApJ...666..694W}). 
Nevertheless these formalisms also rely on the assumption of a non-evolving
color correction and similar conclusions apply.}.  
The coefficients $\alpha$ and $\beta$ are 
unknown parameters which do not depend on redshift.
The color correction
term must account for both dust extinction and any intrinsic color-magnitude
relation with a \emph{unique} coefficent $\beta=\beta_0$.  A redshift-dependent 
color/extinction term will therefore invalidate the assumption that $\beta$ is
constant with $z$, introducing a bias into the parameter estimation.  

The color-based correction makes use of the observable \emph{net} color excess
$c_i$ of each supernova.  In general, this excess can be the sum of several
components,
\begin{equation}
c_i=\sum_k c_{i,k}\;,
\label{eq:color_sum}
\end{equation}
where the $k$-terms describe the intrinsic color, dust reddening due to the
host galaxy,  dust along the line-of-sight and so on. In principle, each 
contribution $c_{i,k}$ can be corrected for using an appropriate 
color-dimming coefficient $\beta_k$: 
\begin{eqnarray}
\delta m_i &=& \sum_k \beta_{i,k}\;c_{i,k}
\label{eq:beta_sum}
\end{eqnarray}
(if attributed to dust extinction, $\beta_k$ corresponds to 
$R_B=A_B/E(B-V)=R_V+1$, following the previously introduced bands). 

Let us now consider the effects of cosmic dust extinction, which we will
denote with the subscript ``$d$''. For simplicity, we will only consider its
redshift dependence, i.e. $\beta_{i,k}=\beta_{\rm d}$ and $c_{i,k}=c_d(z_i)$.  The
brightness of standard candles will appear modified according to
$\delta m_i = \beta_{\rm d}\;c_d(z_i)$. If $\beta_{\rm d}$ differs from the 
best-fit value of $\beta_0$ found in Equation~\ref{eq:chi2}, a 
redshift-dependent magnitude bias
\begin{equation}
\delta m_{{\rm bias}, i} = (\beta_{d}-\beta_0)\,c_d(z_i)\
\label{eq:bias}
\end{equation}
is introduced.  The relation between $\beta_{\rm d}$ and $\beta_0$ depends on the
relative contribution of cosmic dust reddening to the overall color scatter.
Note that such a bias can be either positive or negative.

\begin{figure}
\begin{center}

\includegraphics[width=.5\textwidth]{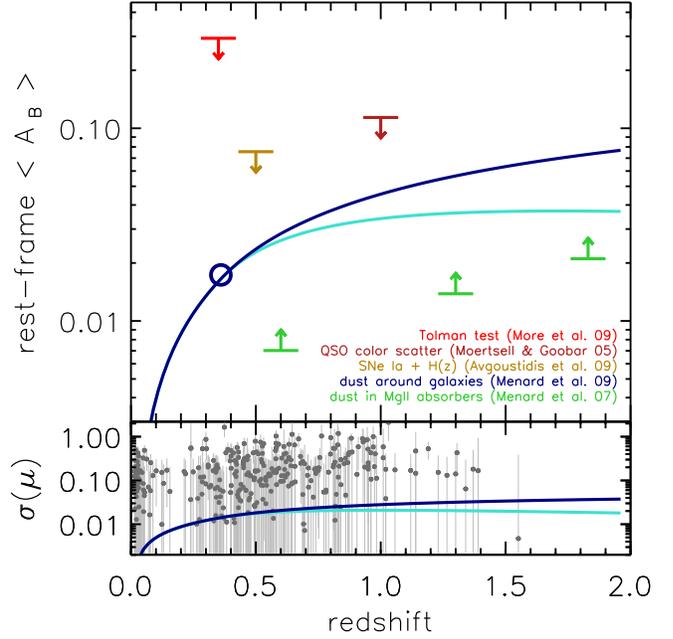}
\caption{\emph{Top:} Opacity of the Universe as a function of redshift, in the
rest-frame $B$ band as estimated by M\'enard et al. (2009).  The light blue 
curve shows an extrapolation of the $z \sim 0.3$ measurements taking into account
the evolution of dust density as a function of redshift. The dark blue curve applies 
a damping term to that extrapolation.
\emph{Bottom:} 
The data points show 
the residuals of the Hubble diagram with 
the Union sample of supernovae.  The blue curves show the expected 
contribution to the scatter in magnitudes due to intergalactic extinction.}
\label{fig:av}

\end{center}
\end{figure}

In order to quantify the amplitude of this potential bias, we use the estimate 
of the opacity of the Universe given by M09. These authors 
reported reddening effects on large-scales around $z\sim0.3$ galaxies, 
characterized by $\beta_{\rm d}=R_{\rm V}+1=4.9\pm2.6$.  Based on these results, 
they proposed several model-dependent estimates of the opacity of the Universe
as a function of redshift.  Here we will consider two cases:
\begin{enumerate}
\item a high-$A_B$ model, shown in Figure~\ref{fig:av} with the dark-blue 
curve. This estimate is calibrated at the redshift of their measurement and 
then extrapolated to higher redshift using an evolution model based on the
observed amount of dust in MgII absorbers.
\item a low-$A_B$ model, where the cosmic dust density for the high-$A_B$ model 
is damped by a factor $(1+z)^{-2}$.  This results in a significant suppression 
of the opacity at $z\gtrsim0.8$. This model, shown with the light-blue
curve, is still consistent with broad observational constraints and only
about a factor $\sim2-3$ higher than the allowed lower limits.
\end{enumerate}

Recent analyses of SNe Ia point toward a value of $\beta_0\simeq2.5$ 
\citep{2008ApJ...686..749K} which is derived from minimizing the residuals in 
the Hubble diagram.  According to Equation~\ref{eq:bias}, a component of 
cosmic dust with  $\beta_{\rm d}=4.9$ will bias the distance modulus estimate by
\begin{eqnarray}
\delta m_{\rm bias}(z=0.5)\simeq 0.01\;{\rm mag}\,.
\end{eqnarray}
Given the simple scaling relations derived in \S\ref{sec:scaling_relations},
this will in turn bias the inferred $\Omega_{\rm M}$ value by
$\delta \Omega_{\rm M}\simeq 0.01$.  This translates into a $\sim3\%$ bias
for $\Omega_{\rm M}\simeq0.3$.  Below we quantify this effect more accurately, 
using existing data.

%- - - - - - - - - - - - - - - - - - - - - - - - - - - - - - - - - - - -      
%- - - - - - - - - - - - - - - - - - - - - - - - - - - - - - - - - - - -      

% =======================================================
\subsection{Application to the Union supernova sample}\label{sec:union_sample}

We now investigate the impact of cosmic dust on a recent supernova dataset:
the ``Union'' sample \citep{2008ApJ...686..749K}.
 These authors have combined
various supernova samples, compiling a ``clean'' dataset of 307 SNe with
$0.015 < z < 1.55$ which they used to infer cosmological parameter constraints.

Using Equation~\ref{eq:chi2}, we sample the parameter space constrained by
this dataset for two cosmological models, using an adaptive importance sampling
algorithm called Population MonteCarlo (for more details, see 
\citealt{2009arXiv0903.0837W})\footnote{In Equation \ref{eq:chi2} the error 
$\sigma_i$ for each supernova contains the covariance of 
$m_{{\rm obs}, i}, s_i, c_i$ (which depends on $\alpha$ and $\beta$),
an intrinsic absolute magnitude scatter of 0.15 and an additional
uncertainty from peculiar velocities of 300 km s$^{-1}$.}.
We first consider a flat $\Lambda$CDM Universe using only constraints from the
supernovae themselves.  This allows us to estimate the constraints on
$\{\Omega_{\rm M}, M, \alpha, \beta\}$, the latter three being the nuisance
parameters intended to absorb the effects of the standard candle normalization.
Our numerical values are presented in Table~\ref{table1} and we show the
posterior distributions of $M$, $\beta$ and $\Omega_{\rm M}$ in
Figure~\ref{fig:posterior} with the gray histograms.  Second, we consider a
dark-energy model with constant equation of state $w$.  In order to obtain
interesting constraints, we add data from the CMB (WMAP5 distance priors;
\citealt{2008arXiv0803.0547K}) and BAO (distance parameter $A$ measured with
the SDSS; \citealt{2005ApJ...633..560E}).  The constraints on the parameters
$\{w, \Omega_{\rm M}, \Omega_{\rm b}, h, M, \alpha, \beta\}$ are given in
Table~\ref{table1} and we show the posterior distribution for
$\Omega_{\rm M}$, $\Omega_{\rm b}$ and $w$ and $h$ in Figure~\ref{fig:posterior2}.
In each case, we recover results similar to \cite{2008ApJ...686..749K}.

% =======================================================
\subsubsection{Smooth dust component}\label{sec:smooth_dust}

%- - - - - - - - - - - - - - - - - - - - - - - - - - - - - - - - - - - -      
\begin{figure}
  \begin{center}
  
    \includegraphics[width=0.45\textwidth]{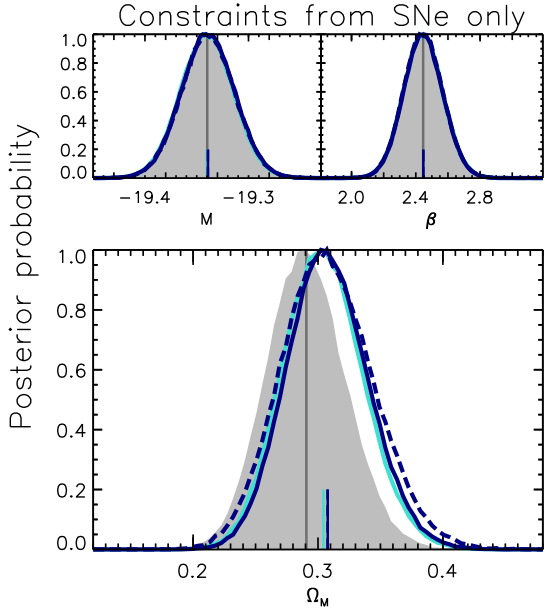}
    \caption{Posterior probabilities for the supernovae absolute magnitude 
      $M$, the color-magnitude scaling coefficient $\beta$ and the matter 
      density $\Omega_{\rm m}$ using the Union sample of supernovae for a 
      $\Lambda$CDM cosmology.  The different models use no dust correction 
      (gray histogram), a correction for the high $A_B(z)$ model 
%      derived by       M09
       (dark blue) and a low $A_B(z)$, for which the dust density is 
      suppressed at high redshifts (light blue).  The dashed curve uses the 
      high $A_B(z)$ and includes the measured error on $\beta_{\rm d}$.   Cosmic 
      dust does not affect the estimate of the nuisance parameters but does 
      impact the constraint on $\Omega_{\rm M}$.}
\label{fig:posterior}
  \end{center}
\end{figure}
  %- - - - - - - - - - - - - - - - - - - - - - - - - - - - - - - - - - - -      

%- - - - - - - - - - - - - - - - - - - - - - - - - - - - - - - - - - - -      
\begin{figure}
  \begin{center}

    \includegraphics[width=0.5\textwidth]{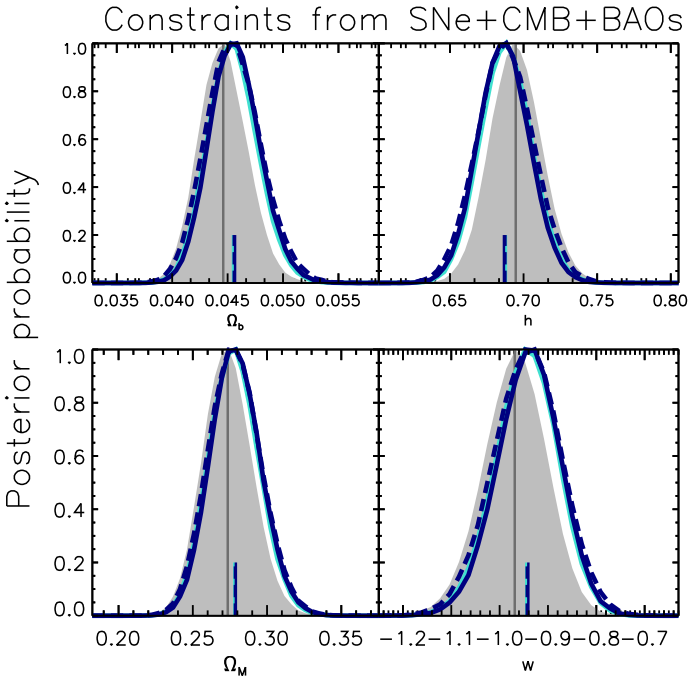}
    \caption{Same as Figure~\ref{fig:posterior} but for a $w$CDM cosmology
      using constraints from SNe Ia+CMB+BAOs.}
    \label{fig:posterior2}
  \end{center}
\end{figure}
%- - - - - - - - - - - - - - - - - - - - - - - - - - - - - - - - - - - -      

After verifying that our method reproduces the current results, we can now 
turn our attention to including the effects of cosmic dust extinction.  First, 
we consider the limiting case of a smooth cosmic dust component.  Using the 
two estimates of the opacity of the Universe introduced above and (for now)
ignoring measurement uncertainties, we can correct the apparent magnitude $m_i$ 
and the net color $c_i$ of each supernova for the expected effect of 
intergalactic dust extinction as a function of supernova redshift.  In 
principle, such corrections should take into account the band used for each 
observation.  In order to simplify our treatment we will however only consider 
the rest-frame $B$-band, which corresponds to the observed $R$-band at the 
average redshift, $\langle z\rangle\simeq0.5$, i.e. the filter used for most of 
the supernovae in this sample.  We then use these corrected magnitudes and 
calculate the constraints on cosmological parameters as done above. 

The results of this calculation are shown in Figure~\ref{fig:posterior} and 
Table \ref{table1}.  We first observe that the \emph{posterior distributions 
on $M$, $\alpha$ and $\beta$ are basically unchanged}.  The color-magnitude 
correction used to convert supernovae into standard candles is therefore not 
sensitive to the presence of a redshift-dependent cosmic extinction.  This is 
expected as a smooth component has a minimal contribution to the scatter in 
magnitude and/or color to which $\beta$ is sensitive.  However, the tilt 
induced in the Hubble diagram by a redshift-dependent cosmic extinction 
significantly affects the estimate of the matter density:
$\delta\Omega_{\rm M}\simeq 0.02$, i.e. a $\sim7\%$ change in amplitude.
This corresponds to a systematic 
shift comparable to the statistical uncertainty.  By comparison, estimates of 
the absolute magnitude $M$ and $\beta$ change by less than $10^{-2}$.  When
extending the cosmological model to a $w$CDM (and adding constraints from CMB 
and BAO), we obtain the results shown in Figure~\ref{fig:posterior2}.  As with 
the $\Lambda$CDM model, the best-fit values of $M$, $\alpha$ and $\beta$ remain unchanged 
(see Table \ref{table1}), while the cosmological parameters are all biased to 
some extent. As shown in the Table, the offsets range from 0.25 to 0.45$\sigma$.

For a more realistic description of the properties of the intergalactic dust
we can include the measurement error on $\beta_{\rm d}$.  To do so, we add a 
Gaussian prior for $\beta_{\rm d}$ to the $\chi^2$ with mean 4.9 and rms $2.6$ 
as indicated by M09.  The results for the high-$A_B(z)$ model are 
shown in Figures~\ref{fig:posterior} \&~\ref{fig:posterior2} using the dark-blue 
dashed lines.  The uncertainty in $\beta_{\rm d}$ propagates into the 
best-fit parameter uncertainties, yielding an increase in the width and 
skewness of the posterior distributions.

These results illustrate that with the current formalism being used to measure 
the luminosity distance from SNe Ia, the color-magnitude correction term does 
not properly account for a redshift-dependent cosmic dust extinction. 
The slope $\beta$ of the color-luminosity relation is not sensitive to such a reddening component.
As a result, the formalism can only correct for a fraction of the extinction effect, as indicated in Eq.~\ref{eq:bias}.
Thus, instead of being absorbed by the nuisance parameters, such a component of dust impacts the 
value of cosmological parameters, resulting in offsets comparable to 
the current statistical errors on $\Omega_{\rm M}$ and $w$.

% ================================================================================================
\begin{table*}
\begin{center}
\begin{tabular}{ll}

\begin{tabular}{rcccc}
\hline
\multirow{2}{*}   				& \multirow{2}{*}{No Correction}& High $A_B$ & High $A_B$   			& Low $A_B$  \\
& & $\beta_{\rm d}=4.9$ &  $\beta_{\rm d}=4.9\pm2.6$ &  $\beta_{\rm d}=4.9$\\
{Parameter}\\

\hline\hline
$\Lambda$CDM: 	$\Omega_{\rm{M}}$   	& $0.291^{+0.032}_{-0.030}$ 	& $0.308^{+0.034}_{-0.031}$~($0.55\sigma$) 	& $0.308^{+0.039}_{-0.035}$~($0.55\sigma$)  	& $0.304^{+0.033}_{-0.031}$~($0.42\sigma$)\\ \hline\hline
$w$CDM: $\Omega_{\rm{b}}$    			& $0.0457^{+0.002}_{-0.002}$ & $0.046^{+0.002}_{-0.002}$~($0.35\sigma$)	& $0.045^{+0.003}_{-0.002}$~($0.25\sigma$)	& $0.045^{+0.002}_{-0.002}$~($0.25\sigma$)  	  \\ \hline
$h$ 				        					& $0.695^{+0.018}_{-0.017}$  	& $0.687^{+0.018}_{-0.017}$~($0.45\sigma$) 	& $0.688^{+0.020}_{-0.019}$~($0.40\sigma$)   & $0.688^{+0.018}_{-0.017}$~($0.40\sigma$)         \\ \hline
$\Omega_{\rm{M}}$   					& $0.273^{+0.017}_{-0.016}$  	& $0.279^{+0.017}_{-0.016}$~($0.36\sigma$) 	& $0.278^{+0.018}_{-0.017}$~($0.30\sigma$) 	& $0.278^{+0.017}_{-0.016}$~($0.30\sigma$)          \\ \hline
$-w$  			        					& $0.968_{-0.061}^{+0.068}$ 	& $0.940_{-0.061}^{+0.067}$~($0.43\sigma$) 	& $0.944_{-0.067}^{+0.072}$(0.37$\sigma$)	& $0.944^{+0.062}_{-0.066}$(0.37$\sigma$)      \\ \hline
\end{tabular}
\begin{tabular}{rc}\hline
Parameter &    \emph{all} models \\\hline\hline
$M$     & $-19.31\pm0.03$\\ 
$\alpha$         & $1.37\pm{0.13}$ \\
$\beta$         & $2.45\pm{0.12}$    \\\hline
\end{tabular}

\end{tabular}

\caption{Constraints on the fitting parameters of Equation~\ref{eq:mu_beta} 
for both a $\Lambda$CDM cosmology (using SNe Ia) and a $w$CDM cosmology (using
SNe Ia+CMB+BAO).  Three estimates of the opacity of
the Universe are considered (see text).  Note that in \emph{all} cases, we obtain the same
best fit values for $\alpha$, $\beta$, $M$.  In other words, the current
parameterization is not sensitive to a component of cosmic dust.  We find that
such a component can bias the estimate of $\Omega_M$ and $w$ by 0.3 to 0.5 $\sigma$ 
with current data.}
\label{table1}
\end{center}
\end{table*}

% =====================================================================================

% =======================================================
\subsubsection{Clumpy dust component}

Since dust is expected to originate from galaxies, the extinction 
effects due to intergalactic dust should be related to the underlying density 
field.  Hence, the dust contribution to the scatter in supernovae 
colors/magnitudes should reflect this correlation.  Using a toy model 
we can investigate the expected impact of this effect on cosmological parameter
estimation.

The number of halos intercepted by an average line-of-sight is given by
\begin{equation}
{N_{\rm g}}= 
\int^z_0\,\sigma\,n\,\frac{c\, (1+z')^{2}}{{H}(z')} {\rm d} z'\,
\label{eq:tau}
\end{equation}
where $\sigma$ and $n$ are the cross-section and number density of the
galaxies responsible for most of the dust extinction.  Assuming that the 
amount of dust is roughly proportional to luminosity and metallicity, M09
claim that most of the dust originates from galaxies with $L\sim L^\star/2$. They 
use $n\simeq 0.037\,h^{3}\,{\rm Mpc}^{-3}$ and $\sigma=\pi\,r_v^2$, with
$r_v\simeq  110\,h^{-1}{\rm kpc}$.

For a Poisson distribution, the level of fluctuation is simply 
$\sim 1/\sqrt{N_{\rm g}}$. An upper limit on the extinction scatter due to intergalactic
dust is then given by ${\rm A_B}(z)/\sqrt{N_g(z)}$. 
This quantity is shown in the lower panel of Figure~\ref{fig:av} for the two models of dust considered
above.
At $z=1$, the average number of halos intercepted by a line-of-sight is of order ten 
and the scatter in $A_B$ is about 0.03 mag. 

Comparing this quantity to the scatter in observed supernova magnitudes from
the Union sample (data points), we see that the the level of scatter due to
cosmic extinction ends up being significantly lower than the observed scatter
in SN magnitudes.  Therefore, this contribution is not expected to
substantially affect the above constraints on cosmological parameters and
leaves our previous conclusions unchanged.

% =======================================================
\section{Discussion and prospects}\label{sec:discussion}

The current formalism used to correct the brightness of supernovae for
color-magnitude trends assumes that all color-magnitude corrections can be 
parametrized by a unique (and therefore redshift independent) coefficient, 
$\beta$.  We have shown that in the presence of several sources of reddening 
this assumption may lead to a biased estimate of the intrinsic brightness of 
SNe and therefore their distance modulus.  In addition, color excesses which 
do not add a significant scatter to the observed SNe Ia brightnesses end up 
affecting the value of cosmological parameters but leave the best-fit color-scaling parameter $\beta$ unchanged.

We have considered the 
effects of the intergalactic dust observed by 
M09 and applied a range of extinction corrections to the Union supernovae sample.
We have shown that the reddening-based correction used in recent works is not sensitive
to such an extinction contribution.
As a result, it only corrects for a fraction of 
the extinction effect, and biases estimates of cosmological parameters.
Considering several scenarios of extinction corrections as a function of redshift we
found biases in $\Omega_M$ and $w$ at a level
ranging from 0.3 to 0.5$\sigma$ of current
statistical errors.

While our analysis indicates a systematic limitation in the current formalism for standardizing SNe Ia,
we remind the reader that the parameter estimates presented above rely on the value of the opacity of the Universe which remains
poorly constrained, especially at high redshift. 
The quantitative results are based on reddening measurements around galaxies at $z\sim0.3$, then extrapolated.
Our analysis is therefore not meant to provide more robust cosmological parameter estimations, but only
\emph{illustrates} the fact that the statistical errors achieved by recent
surveys are becoming comparable to systematic biasing due to extinction. Below 
we suggest a few means by which future surveys can reduce their exposure 
to this particular bias.

\subsection{Minimizing the effect of dust extinction}

Dust extinction effects weaken significantly when considering larger 
wavelengths. One can therefore minimize the amplitude of the bias discussed 
above by observing and estimating supernova magnitudes in redder bands. 
Planned surveys will satisfy this criterion:
\begin{enumerate}
\item JDEM will provide us with observations of supernovae in rest-frame $I$ 
band. For dust with $R_{\rm V}=3$, the extinction effects are expected to be about two
times weaker than those in $B$ band. According to Equation \ref{eq:w}, the 
bias in $w$ will decrease to about 1\%. More detailed calculations applying 
the extinction corrections to simulated JDEM data give about 0.5\% 
(D. Rubin, private communication). 
\item For LSST
\citep{2008arXiv0805.2366I} the reddest filters ($z$ 
and $Y$) should be sufficient for k-correcting to rest-frame $R$ band, which 
should yield roughly a 1.5\% systematic error on $w$ from intergalactic dust using 
the current formalism.
\end{enumerate}

In addition, if the photometry is accurate enough, it becomes possible to correct for reddening effects
object-by-object. With photometry in three or more passbands and a supernova template, it is possible to estimate $R_V$ and $E(B-V)$ for each object. Such an approach no longer requires the assumption of a unique
slope $\beta$ of the supernova color-luminosity relation and might be favored in the future.
As pointed out \cite{2007ApJ...666..694W}, there are currently only 
a few SNe Ia in the literature with the requisite high-precision photometry extending from the rest-frame UV to the near-IR, at $z>0.2$.

\subsection{Detecting and correcting for intergalactic dust extinction with SNe Ia}

Similarly to searches for magnification effects \citep{2005MNRAS.358..101M}, 
the contribution of dust extinction associated with foreground galaxies
can in principle be detected using the supernovae themselves, by 
cross-correlating their observed colors with the density field of foreground 
galaxies:
\begin{equation}
\langle\, \delta c \,\rangle_g(\theta) = 
\langle [c(\phi)-\langle c\rangle] \; \delta_{\rm g}(\phi+\theta) \rangle,
\end{equation}
where
\begin{equation}
\delta_{\rm g}(\theta) \equiv N_{\rm g}(\theta)/\langle N_{\rm g} \rangle\,-1
\end{equation}
and $N_{\rm g}$ is the density of foreground galaxies measured within some
aperture of size $\theta$ around a given supernova.  Note that including
higher-redshift galaxies would not bias the estimate but only add noise.  
Compared to the results presented in this paper, the advantage of such a 
measurement is to directly include the full redshift range from which dust 
extinction effects arise. Similarly, measuring this effect as a function of 
wavelength provides constraints on the coefficient $\beta_{\rm d}$ relevant to 
the considered supernovae sample.

We now estimate the number of supernovae required for such a measurement.
The above correlation is not sensitive to the mean reddening along 
lines-of-sight but only to fluctuations.
The typical reddening excess due to intergalactic dust is given by
$E_{\rm B-V}(z)\;\sigma({N_{\rm g}})$, where $N_{\rm g}$ is defined 
in Equation~\ref{eq:tau}. 
Such a change will be induced to the mean color of supernovae lying behind
the corresponding over/under-densities. The error on the estimate of
the mean color of $N_{\rm S}$ supernovae is given
by $\sigma_{\langle c \rangle} = {\sigma(c)}/{\sqrt{N_{\rm S}}}$, where $\sigma(c)$ 
is the scatter of observed supernova colors. Detecting such a
quantity at the $\nu\,\sigma$-level requires
\begin{equation}
\frac{\nu\,\sigma(c)}{\sqrt{N_{\rm S}}}<E_{\rm B-V}(z)\;\sqrt{N_{\rm g}(z)}\;.
\end{equation}
At $z=0.5$, $\langle E_{\rm B-V} \rangle \simeq5\times10^{-3}$ and
$N_{\rm g}\simeq3$.  For $\sigma(c) \simeq0.1$ (see 
Figure~\ref{fig:av}), we find that $N_{\rm S}\sim 1200$ supernovae are required 
to detect this effect at the 3$\sigma$ level.  At $z=1$, $E_{\rm B-V}\simeq 10^{-2}$ and
$N_{\rm g}\simeq7$, giving $N_{\rm S}\sim130$.  Such detections might 
therefore be within reach, given the expected yields of the next generation of 
supernova surveys.  Hence, the contribution of cosmic dust extinction can be 
corrected for, on average, as done in the present analysis. Note that $N_S\propto\sigma(c)^2$. Therefore, better photometric accuracy would significantly reduce the number of required supernovae to detect the effects of intergalactic reddening.

We are also pointing out that once the mean reddening per galaxy has been estimated,
it is possible to correct the supernova magnitudes object-by-object, 
by subtracting the expected amount of extinction given 
the number of foreground galaxies in the vicinity of each supernova.  If the 
amplitude of intergalactic reddening is properly estimated, the correlation
between foreground galaxy density and \emph{corrected} supernova magnitudes 
should vanish.  This provides a consistency check for the correction of cosmic 
dust.
Such a test emphasizes the necessary interdependence of the large-scale structure and
synoptic aspects of the next-generation surveys.

\section*{Acknowledgements}

We thank David Rubin, Saul Perlmutter, Nao Suzuki, Pierre Astier, Eric Aubourg, Michael Wood-Vasey and Ray Carlberg for useful discussions and
Marek Kowalski for providing us with the Union dataset.

\end{document}